\documentclass[12pt]{article}

\usepackage{scicite}
\usepackage{graphicx}
\usepackage{times}

\usepackage{indentfirst}
\newenvironment{sciabstract}{%
\begin{quote} \bf}
{\end{quote}}

\title{On-demand Integrated Quantum Memory for Polarization Qubits}
\author{

Tian-Xiang Zhu,$^{1, 2}$ Chao Liu,$^{1, 2}$  Ming Jin$^{1, 2}$, Ming-Xu Su$^{1, 2}$, Yu-Ping Liu$^{1, 2}$,\\ Wen-Juan Li,$^{3}$, Yang Ye,$^{3}$ Zong-Quan Zhou$^{1, 2,\ast}$, \\Chuan-Feng Li$^{1, 2, \dag}$, Guang-Can Guo$^{1, 2}$\\
\\
\normalsize{$^{1}$CAS Key Laboratory of Quantum Information,}\\
\normalsize{University of Science and Technology of China, Hefei 230026, China}\\
\normalsize{$^{2}$CAS Center For Excellence in Quantum Information and Quantum Physics,}\\
\normalsize{University of Science and Technology of China, Hefei 230026, China}\\
\normalsize{$^{3}$Center for Micro and Nanoscale Research and Fabrication,}\\
\normalsize{University of Science and Technology of China, Hefei 230026, China}\\
\\
\normalsize{$^\ast$email:zq\_zhou@ustc.edu.cn. $^\dag$email:cfli@ustc.edu.cn.}

}

\date{}

\begin{document}


\baselineskip24pt


\maketitle


\begin{sciabstract}
Photonic polarization qubits are widely used in quantum computation and quantum communication due to the robustness in transmission and the easy qubit manipulation. An integrated quantum memory for polarization qubits is a fundamental building block for large-scale integrated quantum networks.  However, on-demand storing polarization qubits in an integrated quantum memory is a long-standing challenge due to the anisotropic absorption of solids and the polarization-dependent features of microstructures. Here we demonstrate a reliable on-demand quantum memory for polarization qubits, using a depressed-cladding waveguide fabricated in a $\mathrm{^{151}Eu^{3+}}$: $\mathrm{Y_2SiO_5}$ crystal. The site-2 $\mathrm{^{151}Eu^{3+}}$ ions in $\mathrm{Y_2SiO_5}$ crystal provides a near-uniform absorption for arbitrary polarization states and a new pump sequence is developed to prepare a wideband and enhanced absorption profile. A fidelity of ${99.4\pm0.6\%}$ is obtained for the qubit storage process with an input of 0.32 photons per pulse, together with a storage bandwidth of 10 MHz. This reliable integrated quantum memory for polarization qubits reveals the potential for use in the construction of integrated quantum networks.

\end{sciabstract}

% In setting up this template for *Science* papers, we've used both
% the \section* command and the \paragraph* command for topical
% divisions.  Which you use will of course depend on the type of paper
% you're writing.  Review Articles tend to have displayed headings, for
% which \section* is more appropriate; Research Articles, when they have
% formal topical divisions at all, tend to signal them with bold text
% that runs into the paragraph, for which \paragraph* is the right
% choice.  Either way, use the asterisk (*) modifier, as shown, to

\section*{Introduction}

Quantum memories (QMs) are fundamental building blocks for global-scale quantum networks \cite{RN1}. Integrated QMs are particularly useful to meet the requirements of convenience and miniaturization in large-scale applications \cite{RN2}. QMs have been implemented in various physical systems such as single atoms \cite{RN3}, single ions \cite{RN4}, cold atoms \cite{RN5,RN6}, warm atomic vapours \cite{RN7,RN8}, defects in solids \cite{RN9,RN10} and rare-earth-ion doped crystals (REICs) \cite{RN2,RN11}. As a solid-state platform, REICs have extremely-long storage times \cite{RN12,RN13,RN14,RN15}, large bandwidth \cite{RN2,RN16,RN17,RN18,RN19}, large multimode capacities \cite{RN13, RN18, RN20,RN21,RN22,RN23,RN24,RN25,RN26} and can be easily manufactured to be integrated QMs \cite{RN2, RN16, RN17,RN26,RN27,RN28,RN29,RN30,RN31,RN32}.

As the natural carrier of information in quantum networks, photonic qubits can be coded into various degrees of freedom, such as polarization, time, path and frequency. Among them, the polarization qubits are widely adopted due to the easy manipulation with waveplates and easy transmission in a single temporal and spatial mode. QMs for photonic polarization qubits have been demonstrated in REICs \cite{RN33,RN34,RN35,RN36,RN37}, single atoms \cite{RN3}, cold atomic ensembles \cite{RN5, RN38, RN39} and warm atomic vapours \cite{RN40}. However, on-demand integrated QMs for polarization qubits have not yet been achieved. There are two problems need to be solved. First, unlike free-space atoms \cite{RN3}, most solids have anisotropic absorption. For REICs, pioneering works solved this problem by storing two different polarization components into spatially-different parts of the medium \cite{RN33,RN34,RN35,RN37}. Such approaches are not suitable for integrated QMs due to the complex optical setup which may introduce unwanted losses. An amorphous glass fiber can provide a uniform absorption \cite{RN36} at a price of much less storage times as compared to that of single-crystal hosts. Second, to date, all integrated QMs fabricated on REICs are polarization dependent \cite{RN2,RN12,RN17,RN26,RN27,RN28,RN29,RN30,RN31,RN32, RN41,RN42,RN43,RN44,RN45,RN46,RN47,RN48,RN49,RN50,RN51}, i.e. only one polarization mode is supported in these micro/nano structures. 

Here, we demonstrate an integrated QM for polarization qubits by fabricating polarization-independent waveguides in a uniformly-absorbing single crystal (site-2 $\mathrm{^{151}Eu^{3+}}$ ions in $\mathrm{Y_2SiO_5}$ crystals). The depressed-cladding optical waveguide, also known as the type-III waveguide \cite{RN52}, is fabricated using femtosecond laser micromachining (FLM). This optical waveguide is further combined with on-chip electrodes to facilitate on-demand quantum storage based on the Stark-modulated atomic frequency comb protocol \cite{RN31, RN53, RN54}.

\section*{Result}

\section*{Device and Experiment setup}
The substrate is a $0.1\%$ doped $\mathrm{^{151}Eu^{3+}}$:$\mathrm{Y_2SiO_5}$ crystals with a size of $3 \times 12 \times 4 \mathrm{~mm}^{3}$ along $D1 \times D 2 \times b$ axes. The type-III waveguide is fabricated by a FLM system (WOPhotonics). The details of the fabrication process are provided in the Method. In the type-III waveguide, the low refractive-index optical barriers are separated by a few microns and the optical field is confined among them \cite{RN52,RN55}. To achieve a near-uniform absorption for arbitrary polarization states, we use $Eu^{3+}$ ions that locates at site-2 of $\mathrm{Y_2SiO_5}$ crystals and the light propagates along the D2 direction \cite{RN56}.

For convenience, we denote the polarization state parallel to the $D1(b)$ axis as $|H\rangle(|V\rangle)$. As shown in Fig. 1, the fabricated waveguide has a guide mode of 9.0 $\mu m$ $\times$ 10.3 $\mu m$ $(D1 \times b)$. The insertion losses (including coupling losses and propagation losses) are 0.85 (0.84) dB for $|H\rangle(|V\rangle)$ and the coupling efficiency of the output mode to a single-mode fiber (SMF) is approximately $80\%$. The optical waveguide is fabricated at a depth of 20 um beneath the surface of the crystal for efficient interface with the on-chip electrodes. The two parallel electrodes have a width of 200 um and a spacing of 100 um. An arbitrary function generator (Tektronix, AFG3102C) is employed to directly drive the electrodes to introduce the required electric field on the $\mathrm{^{151}Eu^{3+}}$ ions inside the optical waveguide.

The schematic drawing of the experimental setup is presented Fig. 1 (A). The 580-nm laser is a frequency-doubled semiconductor laser with a stabilized linewidth of below 1 kHz. Both the signal beam and the preparation beam are controlled by acousto-optic modulators (AOMs) which are driven by arbitrary waveform generators. These two beams are combined into a single beam and coupled into the waveguide memory which is cooled down to 3.4 K using a cryostat (Montana Instruments). The polarization state of the signal is prepared and analyzed with standard waveplates and polarization beam splitters. The signal beam is decreased to single-photon levels and two shutters are employed to protect the single photon detectors from the strong preparation beam.

\section*{The Stark effect for site-2 $\mathrm{^{151}Eu^{3+}}$ ions in $\mathrm{Y_2SiO_5}$}

Atomic frequency comb (AFC) is a well-established quantum storage protocol in REICs with advantages of large bandwidth \cite{RN2, RN16, RN18, RN19}, high fidelity \cite{RN11, RN21, RN33,RN34,RN35}, and excellent multi-mode capability \cite{RN13, RN18, RN20,RN21,RN22,RN23,RN24,RN25,RN26}. The recently-proposed Stark-modulated atomic frequency comb (SMAFC) protocol \cite{RN31, RN53, RN54} could further enable on-demand readout by employing electric field pulses to actively control the rephasing process of AFC.

In $\mathrm{Y_2SiO_5}$ crystals, when the electric fields are parallel to the b axis or in the mirror plane, the doped ions will be split into two groups by the Stark effect \cite{RN31, RN53, RN54}.  These two groups of ions have frequency shifts of the equal amounts but with opposite directions. The linear Stark splitting of the $\mathrm{Eu^{3+}}$ ions at two crystalline sites in $\mathrm{Y_2SiO_5}$ crystals have been measured when the electric fields are parallel to the $D1$ and $D2$ axes \cite{RN57}. Here, we further characterize the Stark splitting of site-2 $Eu^{3+}$ ions when the electric field is parallel to the b axis.

The Stark splitting are observed using the spectral hole burning techniques \cite{RN31, RN57}. Under a biased electric field along with the b axis, an antihole will split into two antiholes. As shown in Fig. 2 (C), this experiment consists of four steps. The first step is to initialize the experiment through a large-scale sweeping pump at the center frequency $f_{0}$  with a bandwidth of 125 MHz. The second step is the burning of spectral pit at the center frequency $f_{0}$  with a bandwidth of 6 MHz. The third step is the preparation of an antihole. There are many possible choices of pump frequencies for this step and here we pump at the frequency of $f_{0}+57.24$ MHz and  $f_{0}-57.24$ MHz, using weak gaussian pulses with duration of 4 ms to get a narrow antihole at $f_{0}$. Finally, we measure the splitting of the antihole with the applied electric field along b axis. As shown in Fig. 2 (A) and Fig. 2 (B), the linear Stark splitting along the b axis of site-2 $\mathrm{^{151}Eu^{3+}}$ are $5.66 \pm 0.03 \mathrm{KHz} /\left(\mathrm{V} \cdot \mathrm{cm}^{-1}\right)$ and $-5.71 \pm 0.04 \mathrm{KHz} /\left(\mathrm{V} \cdot \mathrm{cm}^{-1}\right)$ with the average value is $5.69 \pm 0.04 \mathrm{KHz} /\left(\mathrm{V} \cdot \mathrm{cm}^{-1}\right)$.

\section*{On-demand storage of weak-coherent pulses}

Then we continue our experiment with on-demand storage based on SMAFC. Considering combs with Gaussian line shapes, the storage efficiency of the SMAFC memory is \cite{RN53}
\begin{equation}
       \eta=\frac{\pi}{4 \ln (2)}\left(\frac{d}{F}\right)^{2} \cdot \exp \left(-\sqrt{\frac{\pi}{4 \ln (2)}} \frac{d}{F}\right) \cdot \exp \left(-\frac{\pi^{2}}{2 \ln (2)} \frac{m^{2}}{F^{2}}\right)
    ,
\end{equation}
where d is the absorption depth, F is the finesse of the comb, and m is the order of the SMAFC echo. It is obvious that high finesse (F) and high absorption depth (d) are extremely important for achieving high efficiency storage. However, the ${ }^{7} \mathrm{~F}_{0} \rightarrow{ }^{5} \mathrm{D}_{0}$ transition of $\mathrm{Eu^{3+}}$ is only weakly allowed in $\mathrm{Y_2SiO_5}$ crystals and this problem is even worse when working with site-2 $Eu^{3+}$ ions \cite{RN56}. In our experiment, the natural sample absorption d =1.46 (1.64) for $|H\rangle(|V\rangle)$. To this end, we develop a pump sequence to prepare an enhanced and flat absorption profile with site-2 $Eu^{3+}$ ions. Successively utilizing the pump light at the center frequency of  $f_{0}+33.02$ MHz and  $f_{0}-33.02$ MHz with the chirping bandwidth of 46 MHz, we prepare a 12.68-MHz enhanced absorption peak at the center frequency of $f_{0}$ . Using our optical pumping scheme, the enhanced d is 3.85 \cite{RN3,RN35} for $|H\rangle(|V\rangle)$ with the average magnitude of enhancement of 2.64 times as compared to the original absorption. Detailed analysis of this pump sequence is provided in the Method section.

Next, we create an AFC structure with a comb spacing of 2 MHz and a total bandwidth of 10 MHz by parallel comb preparation scheme \cite{RN58}. The comb structures are shown in Fig. 3 (A) and Fig. 3 (B) for input light of two orthogonal polarization states. Our AFC is an imperfect square comb and the expected AFC storage efficiency is between that can be obtained with the Gaussian comb (Eq. (1)) and the square comb \cite{RN58}. The measured efficiencies of the 500-ns AFC are $23.3\pm0.3\%$ ($26.9\pm0.3\%$) for $|H\rangle(|V\rangle)$. The theoretical efficiencies are $29.6\%$ ($32.0\%$) for $|H\rangle(|V\rangle)$ for the square-comb model and $21.5\% (24.2\%)$ for the Gaussian-comb model. 

The simple two-level AFC only allows predetermined storage times. To actively control the AFC rephasing process, we further applied electric field pulses to implement SMAFC. The first electric pulse with a duration of 85 ns and a voltage of 5 V is applied after the $1_{st}$ AFC echo, to prevent the emission of $1_{st}$ to $(n-1)_{th}$ order AFC echoes. Then the second pulse with reversed polarity is applied before the nth order AFC echo, so that to compensate the Stark-induced phase and readout the nth order AFC echo in an on-demand fashion. Photon counting histograms for input states of  $|H\rangle(|V\rangle)$ are presented in Fig. 3 (C) and (D), respectively. The input pulses contain 0.32 photons on average in the input side of the cryostat. The internal storage efficiencies of the 1-$\mu$s SMAFC echo are $13.2\pm0.8\%$ and $15.5\pm0.8\%$ for $|H\rangle$ and $|V\rangle$, respectively. Since the efficiencies of other optical components, such as the single-mode fiber coupling, also have slight polarization dependence, the device efficiencies (between the input of the cryostat and the detector) for 1-$\mu$s SMAFC storage are $7.0\pm0.4\%$ and $7.6\pm0.4\%$ for for $|H\rangle$ and $|V\rangle$, respectively, which is approximately balanced for storage of arbitrary polarization states (Fig. 3 (E)).

\section*{On demand storage of polarization qubits}

With an average photon number of 0.32 photons per input pulse,  the measured signal to noise ratios (SNRs) are $1265\pm729$ and $1039\pm520$ for $|H\rangle(|V\rangle)$, respectively. Such high SNR could allow high-fidelity qubit storage.  Quantum process tomography (QPT) is performed to benchmark the performance of qubit storage of our device \cite{RN24}. The process matrix is constructed based on the maximum likelihood estimation, with a fidelity of $99.4\pm0.6\%$ to the ideal identical storage process (Fig. 4). Here the error bar refers to one standard deviation given by the Monte Carlo simulation assuming Poisson statistics of photons. Taking into account the finite storage efficiency of $7.0\%$ and the Poisson statistics of the weak coherent pulse, the maximal achievable fidelity using a classical measure and prepare strategy is $76.2\%$ \cite{RN34}. Our result outperforms such strict classical bound by 39 standard deviations, unambiguously demonstrating that this device operates in the quantum regime.

\section*{Discussion} 
 We fabricate a depressed-cladding optical waveguide and on-chip electrodes in a $\mathrm{^{151}Eu^{3+}}$: $\mathrm{Y_2SiO_5}$ crystal. On-demand storage of polarization qubits with a bandwidth of 10 MHz is achieved by implementing SMAFC protocol using site-2 $\mathrm{Eu^{3+}}$ ions. The fidelity of the storage process reaches $99.4\pm0.6\%$, which is approximately the same as that achieved with the bulk material \cite{RN21} but provides the additional capability of on-demand retrieval.
 
The unique advantage of the current integrated QM is the capability of supporting arbitrary polarization which could play an essential role in quantum networks involving polarization qubits. A straight forward extension of the current work would be the implementation of spin-wave AFC storage which can provide extended storage time up to 1 hour using the current material \cite{RN15}. In such case, strong control pulses are required and our device could further provide high-contrast polarization filtering of strong control pulses which co-propagate with single-photon signal field in the same waveguide. Moreover, such integrated structure could enable high-density spatial-multiplexed storage for transportable quantum memories \cite{RN15, RN59} and quantum repeaters \cite{RN1, RN18} which are viable solutions for long-distance quantum communication.

\section*{Materials and Methods}

\section*{The device fabrication}

The fabrication of the type-III optical waveguides is accomplished by a commercial FLM system (WOPhotonics). Since this waveguide need to support arbitrary polarization, the cross-section of the type-III waveguide is designed to be a circle composed of several tracks. Here, 20 tracks are fabricated to form the waveguide. In the manufacturing process, we set the femtosecond-laser in the wavelength of 1030 nm, the repetition rate of 201.9 kHz, the pulse duration of 210 fs and a per-pulse energy of 64 nJ. The laser pulses polarized along the D2 axis are focused by a X100 objective with a numerical aperture is 0.7 and irradiate the crystal along the D1 axis. The laser beam moves along the D2 axis at a speed of 1mm/s. To reduce the aberration of the laser pulses and for efficient interface with on-chip electrodes, we set the center of the type-III waveguides to a depth of 20 um beneath the surface of the crystal with diameter of 21 um. The two electrodes are fabricated using ultraviolet lithography (Karl Suss, MABA6) and electron beam evaporation (K.J. Lesker, LAB 18).

\section*{Preparation of the enhanced absorption profile}

Based on the level structure shown in Fig. 1(B), Fig. 5 further provides the aligned energy diagram of 9 classes of site-2 $\mathrm{^{151}Eu^{3+}}$ ions. The zero detuning represents the frequency f of the incident light and causes nine different classes of transitions. For example, for class-I ions, if we burn a hole at zero detuning, their $|1 / 2\rangle_{g}$ state will be empty and all population is distributed in $|3 / 2\rangle_{g}$ and $|5 / 2\rangle_{g}$ ground states.  Antiholes in the frequencies corresponding to the $|3 / 2\rangle_{g}$ state (indicated with red lines) and $|5 / 2\rangle_{g}$ state (indicated with green lines) would be observed. If we burn a broadband spectral pit rather than a single-frequency hole, the broadband antiholes will also be formed at the corresponding frequencies. Here, for simplicity, we consider that the lines widening to the right. The useful bandwidth for each class of ions is limited by the spacing between the neighboring transition on the left and the transition at the frequency of $f$.  For example, for class-VIII ions, the useful bandwidth would be limited to 5.12 MHz, which is the spacing between the red line on the left and the green line with zero detuning at the frequency of $f$. 

The first pump sequence is in the frequency range of [0,\ +46\ MHz] (denoted as pump-1). We can analyze its impact on the target absorption band in the frequency range of [+49.68\ MHz,\ +62.36\ MHz] through a detailed analysis of all nine classes of ions. 

Taking the class-V ions for example, the useful bandwidth is [0,\ +32.82\ MHz]. Class-V ions, as defined by those ions with the $|3 / 2\rangle_{\mathrm{g}} \rightarrow|3 / 2\rangle_{\mathrm{e}}$ transition in the frequency range of [0,\ +32.82\ MHz], are all pumped away from $|3 / 2\rangle_{g}$ state. In particular, for those class-V ions with the $|3 / 2\rangle_{\mathrm{g}} \rightarrow|3 / 2\rangle_{\mathrm{e}}$ transition in the frequency range of [0,\ +16.46\ MHz], they are further pumped away form $|1/ 2\rangle_{g}$ state  because their $|1 / 2\rangle_{\mathrm{g}} \rightarrow|3 / 2\rangle_{\mathrm{e}}$ transition is also addressed by the pump-1 sequence, as indicated by the blue line on the right of zero detuning.

Considering those class-V ions with the $|3 / 2\rangle_{\mathrm{g}} \rightarrow|3 / 2\rangle_{\mathrm{e}}$ transition in the frequency range of [0,\ +12.68\ MHz], they are completely polarized to $|5 / 2\rangle_{g}$  state and contribute to the target absorption band at [+49.68 MHz, +62.36 MHz] through $|5 / 2\rangle_{\mathrm{g}} \rightarrow|1 / 2\rangle_{\mathrm{e}}$ transition.

Considering those class-V ions with the $|3 / 2\rangle_{\mathrm{g}} \rightarrow|3 / 2\rangle_{\mathrm{e}}$ transition in the frequency range of [+20.14\ MHz,\ \ +32.82\ MHz], their population are distributed in $|1 / 2\rangle_{g}$ and $|5 / 2\rangle_{g}$  ground states and partially contribute to the target absorption band at [+49.68 MHz, +62.36 MHz] through transition $|1 / 2\rangle_{\mathrm{g}} \rightarrow|3 / 2\rangle_{\mathrm{e}}$. 

In summary, class-V ions with the $|3 / 2\rangle_{\mathrm{g}} \rightarrow|3 / 2\rangle_{\mathrm{e}}$ transition in the frequency range of [0,\ +16.46\ MHz] and [+20.14\ MHz,\ \ +32.80\ MHz] both contribute the target absorption band in the frequency range of [+49.68\ MHz,\ +62.36\ MHz]. 

After analyzing other classes of ions using the same method, we find that the Class-IV, -V and -VI ions contribute the target absorption band in the frequency range of [+49.68\ MHz,\ +62.36\ MHz]. Class-III ions contribute the absorption in the frequency range of [+49.68\ MHz,\ +57.24\ MHz] and the Class-VII, -VIII and -IX ions contribute the absorption in the frequency range of [+57.24\ MHz,\ +62.36\ MHz]. As a result, the absorption in the frequency range of [+49.68\ MHz,\ +57.24\ MHz] is weaker than that in the frequency range of [+57.24\ MHz,\ +62.36\ MHz].

To obtain a flat absorption band, we further apply a symmetrical pump (denoted as pump-2) in the frequency range of [+66.04,\ +112.04\ MHz]. This is equivalent to pump at [-46\ MHz,\ 0] and to analyze its impact on the target absorption band in the frequency range of [-62.36\ MHz,\ \ -49.68\ MHz]. For simplicity in analysis, here we can assume the lines widening to the left. Pump-2 can perfectly compensate the imbalanced absorption band produced from pump-1. 

In summary, after applications of the pump-1 and pump-2 in sequence, we get a flat enhanced absorption profile with a bandwidth of 12.68 MHz which allows wideband AFC storage. The prepared AFC as shown in Fig. 3 have nearly identical absorption peaks, demonstrating the success of the pumping strategy.

% Your references go at the end of the main text, and before the
% figures.  For this document we've used BibTeX, the .bib file
% scibib.bib, and the .bst file Science.bst.  The package scicite.sty
% was included to format the reference numbers according to *Science*
% style.

%BibTeX users: After compilation, comment out the following two lines and paste in
% the generated .bbl file. 

\bibliography{scifile}

\bibliographystyle{Science}

\section*{Acknowledgments}

Funding: This work is supported by the National Key R\&D Program of China (No. 2017YF A0304100), the National Natural Science Foundation of China (Nos. 11774331, 11774335, 11821404 and 11654002), the Fundamental Research Funds for the Central Universities (No. WK2470000026 and No. WK2470000029) and this work was partially carried out at the USTC Center for Micro and Nanoscale Research and Fabrication. Z.-Q.Z acknowledges the support from the Youth Innovation Promotion Association CAS.

Author contributions: Z.-Q.Z. designed the experiment; T.-X. Z. fabricated the device. T.-X. Z. performed the experiment and analyzed the data with the help from C. L., M. J., Y.-P. Liu and M.-X. S.; W.-J. L. and Y. Y. contributed to the fabrication of electrodes; T.-X. Z. and Z.-Q.Z. wrote the manuscript; Z.-Q.Z. and C.-F.L. supervised the project. All authors discussed the experimental procedures and results.

Competing interests: All other authors declare they have no competing interests.

Data and materials availability: All data are available in the main text or the supplementary materials. Additional data related to this paper may be requested from the authors.

\clearpage

\section*{Figures and Tables}

\includegraphics[width=1\linewidth]{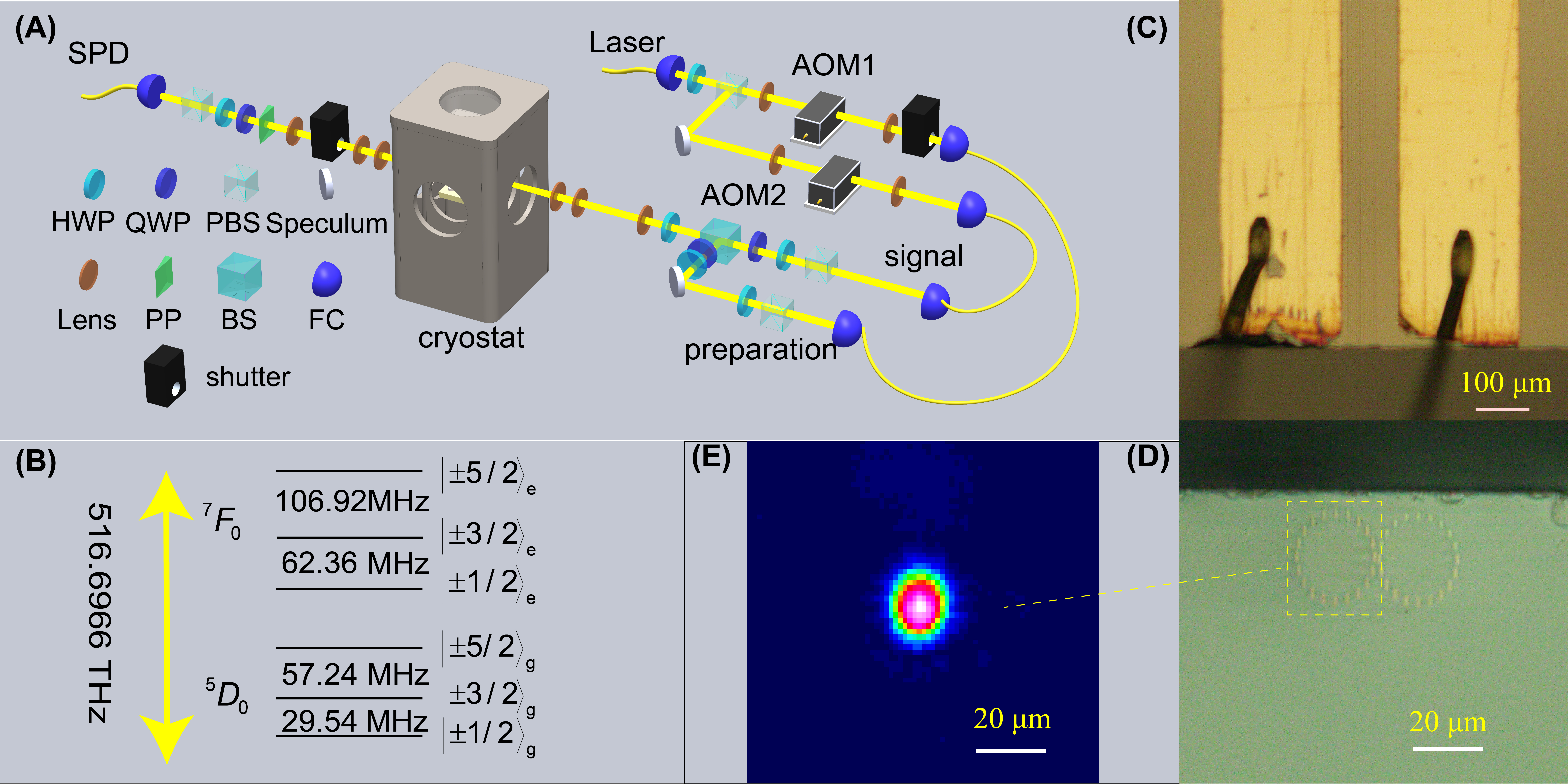}
\noindent {\bf Fig. 1. Experimental setup.} {\bf(A)} Schematic of the setup. The signal and the preparation beams generated by the acousto-optic modulation (AOMs) in double-pass configurations are combined by a beam splitter (BS) and coupled into the type-III waveguide by lens group. The polarization of the signal qubits is prepared by a polarization beam splitter (PBS), a half wave plate (HWP) and a quarter-wave plate (QWP). The HWP after the BS is employed to match the axial direction of the crystal. Another lens group is used for collecting the emergent light from the waveguide. A PP is used to compensate the phase difference accumulated in the setup and the polarization of signal photons is analyzed using QWP, HWP and PBS. The signal is finally coupled into a fiber coupler (FC) and detected by a single photon detector (SPD). {\bf(B)} Energy diagram of  transition for site-2 $\mathrm{^{151}Eu^{3+}}$ ions in $\mathrm{^{151}Eu^{3+}}$: $\mathrm{Y_2SiO_5}$ at zero magnetic field \cite{RN60}. {\bf(C)} The vertical view of type-III waveguides and the electrodes with a scale bar is 100 $\mu$m. {\bf(D)} The front view of the two type-III waveguides with a scale bar is 20 $\mu$m. The left one is the one used in the experiment. {\bf(E)} The beam profile at the exit of the waveguide as measured by a charge coupled device with a scale bar is 20 $\mu$m.

\clearpage
\includegraphics[width=1\linewidth]{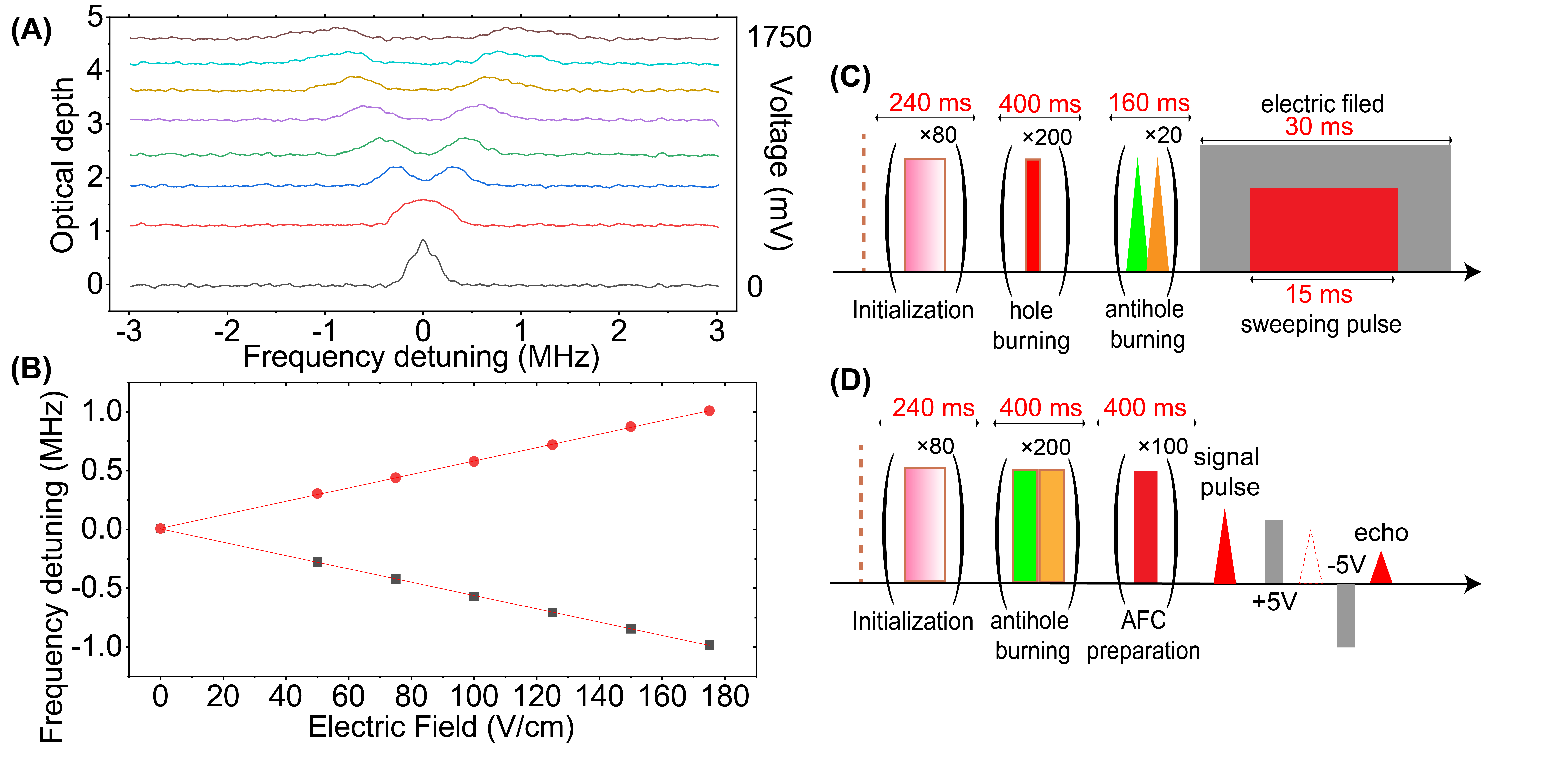}
\noindent {\bf Fig. 2. The linear Stark splitting for site-2 $\mathrm{^{151}Eu^{3+}}$ ions in $\mathrm{Y_2SiO_5}$ crystals.} {\bf(A)} The measured spectral antiholes with variable applied voltages along crystal b-axis. From bottom (black) to top, the voltage increases from 0 mV to 1750 mV with a step of 250 mV. {\bf(B)} The frequency detuning as a function of bias electric fields. The two group ions have Stark shifts of similar magnitudes but opposite directions. The average Stark splitting is $5.69 \pm 0.04 \mathrm{KHz} /\left(\mathrm{V} \cdot \mathrm{cm}^{-1}\right)$ based on linear fittings of the data. {\bf(C)} Time sequence for the measurement of the linear Stark splitting. After preparation of the antihole, the spectral feature is measured with sweeping optical pulse with applied electric fields. {\bf(D)} Time sequence for SMAFC storage. After preparation of AFC, two electric pulse actively control the readout times of the echo. Standard AFC echo is suppressed by the first electric pulse and the SMAFC echo is retrieved after the second electric pulse.

\clearpage
\includegraphics[width=1\linewidth]{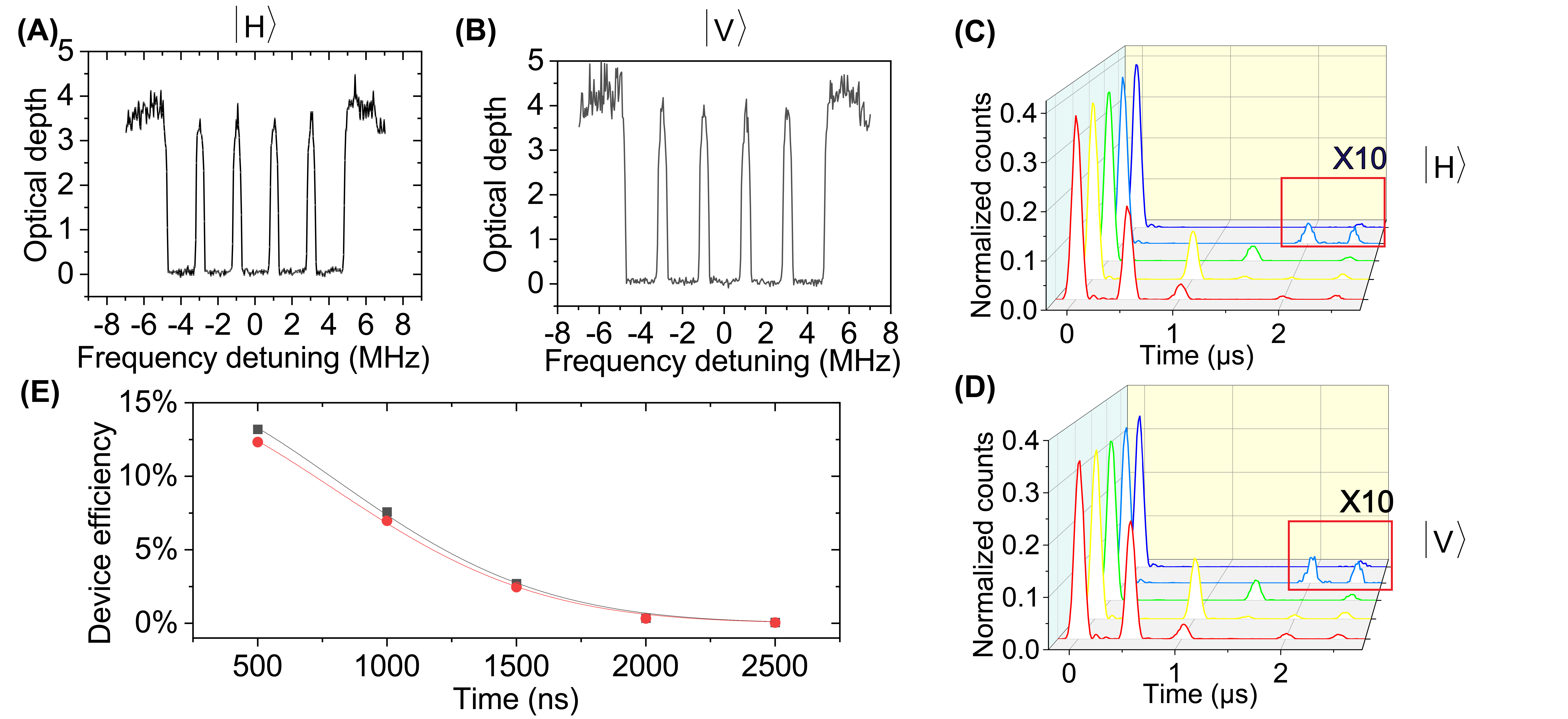}
\noindent {\bf Fig. 3. SMAFC memory for weak coherent pulses.} {\bf(A)} and {\bf(B)} are the measured comb structures with input states of $|H\rangle$ and $|V\rangle$, respectively. The average peak absorption depth is 3.6 (4.0) for $|H\rangle(|V\rangle)$ with negligible background absorption. (C) and (D) are photon counting histograms for SMAFC storage, with input states of $|H\rangle$ and $|V\rangle$, respectively. Echoes retrieved after 2 µs are magnified by 10 times. (E) The device efficiency of the SMAFC memory for input states of $|H\rangle$ (red) and $|V\rangle$ (black). According to Eq. 1, the fitted finesse of the combs is $6.0 \pm 0.1(6.0 \pm 0.1)$ for $|H\rangle(|V\rangle)$. 

\clearpage
\includegraphics[width=1\linewidth]{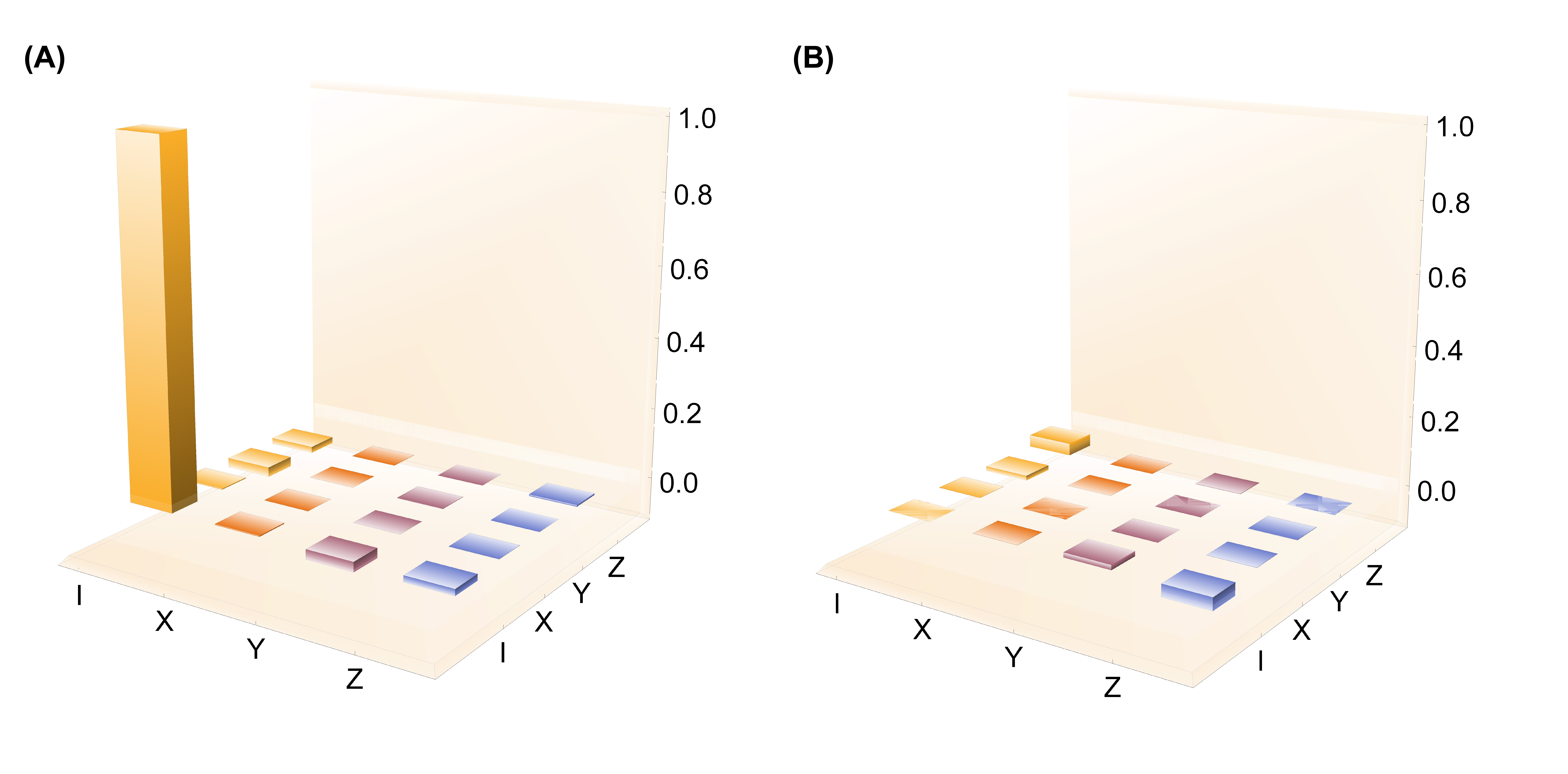}
\noindent {\bf Fig. 4. The reconstructed process matrix of the storage process.} {\bf(A)} The real part of the process matrix as obtained by the quantum process tomography \cite{RN61}. The horizontal axes are the basis operators in the two-dimensional Hilbert space. {\bf(B)} The imaginary part of the process matrix with the largest imaginary element of 0.035.
\clearpage
\includegraphics[width=1\linewidth]{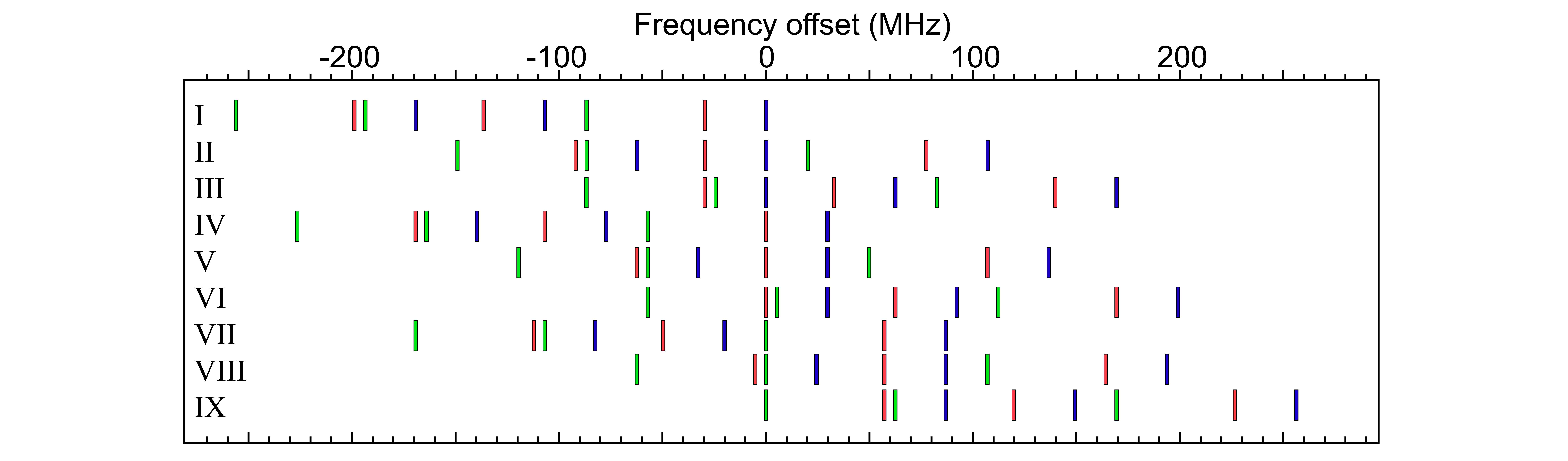}
\noindent {\bf Fig. 5. Aligned energy diagram of 9 classes of site-2 151Eu3+ ions.} The zero detuning represents the frequency f of the incident light. The blue, red and green lines represent the transitions related to the ground states $|1 / 2\rangle_{g}$, $|3 / 2\rangle_{g}$ and $|5 / 2\rangle_{g}$ respectively. For instance, ions with the frequency $f$ corresponding to the transition $|1 / 2\rangle_{g} \rightarrow|5 / 2\rangle_{e}$ are class-I ions, and ions with the frequency $f$ corresponding to the transition $|5 / 2\rangle_{g} \rightarrow|1 / 2\rangle_{e}$ are class-IX ions.

\end{document}